\documentclass[twocolumn, showpacs,preprintnumbers,amsmath,amssymb]{revtex4}
\usepackage{enumitem}
\usepackage{amssymb}
\usepackage{xcolor}
\usepackage{graphicx}
\usepackage{dcolumn}
\usepackage{bm}
\usepackage{multirow}
\usepackage{booktabs}
\usepackage{longtable}
\usepackage{natbib}
\usepackage{lscape}
\usepackage{soul}
\usepackage{amssymb}
\usepackage{amsmath}	
\usepackage{color}
\usepackage{cases} 

\usepackage{array}
\newcommand{\Kmat}{\mbox{\boldmath $\mathcal{K}$}}

\newcommand{\Vmat}{\mbox{\boldmath $\mathcal{V}$}}

\newcommand{\Umat}{\mbox{\boldmath $\mathcal{U}$}}
\newcommand{\etamat}{\mbox{\boldmath $\eta$}}
\newcommand{\numat}{\mbox{\boldmath $\nu$}}
\newcommand{\Cmat}{\mbox{\boldmath $\cal C$}}
\newcommand{\Smat}{\mbox{\boldmath $\cal S$}}
\newcommand{\SSmat}{\mbox{ $S$}}
\newcommand{\Xmat}{\mbox{\boldmath $X$}}

\begin{document}
\preprint{APS/123-QED}

\title{Dissociative recombination, and vibrational excitation of CO$^{+}$: model calculations and comparison with experiment}
\author{J. Zs Mezei$^{1,2,3}$}
\author{R. D. Backodissa-Kiminou$^{1}$}
\author{ D. E. Tudorache$^{1,4}$}
\author{V. Morel$^{5}$}
\author{K. Chakrabarti$^{1,6}$}
\author{O. Motapon$^{7}$}
\author{O. Dulieu$^{2}$}
\author{J. Robert$^{2}$}
\author{W.-\"U. L. Tchang-Brillet$^{8}$}
\author{A. Bultel$^{5}$}
\author{X. Urbain$^{9}$}
\author{J. Tennyson$^{10}$}
\author{K. Hassouni$^{11}$}
\author{I. F. Schneider$^{1,2}$}\email[]{ioan.schneider@univ-lehavre.fr}
\affiliation{$^{1}$LOMC CNRS$-$Universit{\'{e}} du Havre$-$Normandie Universit{\'{e}}, 76058 Le Havre, France}
\affiliation{$^{2}$LAC, CNRS$-$Universit\'e Paris-Sud$-$ENS Cachan$-$Universit\'e Paris-Saclay, 91405 Orsay, France}%
\affiliation{$^{3}$HUN-REN Institute for Nuclear Research (ATOMKI), H-4001 Debrecen, Hungary}%
\affiliation{$^{4}$Lab. EM2C, CNRS-\'{E}cole Centrale Paris, 92295 Ch\^{a}tenay-Malabry, France}%
\affiliation{$^{5}$CORIA CNRS$-$Universit\'{e} de Rouen$-$Universit{\'{e}} Normandie, F-76801 Saint-Etienne du Rouvray, France}%
\affiliation{$^{6}$Dept. of Mathematics, Scottish Church College, Calcutta 700 006, India}
\affiliation{$^{7}$LPF, UFD Math., Info. Appliq. Phys. Fondamentale, University of Douala, P. O. Box 24157, Douala, Cameroon}%
\affiliation{$^{8}$LERMA, Observatoire de Paris, CNRS-Sorbonne Universit\'es-UPMC  Univ Paris 06, F-92195 Meudon, France}%
\affiliation{$^{9}$Institute of Condensed Matter and Nanosciences, PAMO - Louvain-la-Neuve, 1348, Belgium}
\affiliation{$^{10}$Dept. of Physics and Astronomy, University College London, WC1E 6BT London, UK}%
\affiliation{$^{11}$LSPM, CNRS$-$Universit\'e Paris 13$-$USPC, 93430 Villetaneuse, France}%
\date{\today}

\begin{abstract}
The latest molecular data - potential energy curves and Rydberg$/$valence interactions - characterizing the super-excited electronic states of CO are reviewed, in order to provide inputs for the study of their fragmentation dynamics. Starting from this input, the main paths and mechanisms for CO$^+$ dissociative recombination are analyzed; its cross sections are computed using a method based on Multichannel Quantum Defect Theory. Convoluted cross sections, giving both isotropic and anisotropic Maxwellian rate-coefficients, are compared with merged-beam and storage-ring experimental results. The calculated cross sections underestimate the measured ones by a factor of $2$, but display a very similar resonant shape. These facts confirm the quality of our approach for the dynamics, and call for more accurate and more extensive molecular structure calculations.
\end{abstract}

\pacs{33.80. -b, 42.50. Hz}

\maketitle

\section{Introduction}\label{sec:intro}

Carbon monoxide (CO) is the most abundant molecule detected in the interstellar
medium after molecular hydrogen, 
 and being heteronuclear is much easier to be detected in emission. 
Its observation at long wavelengths, as in radio astronomy, is the source of
the much of the information obtained on different interstellar
environments, from molecular clouds in which protostars form and irradiate the
surrounding residual material, to the circumstellar disks that surround stars at
the end of their lives  and to the cometary gases. 
Its spectrum has been intensively studied in astrophysical context, since its
large binding energy implies that many electronically excited states possess
discrete
line spectra, see~\cite{rosen98,chakrabarti06,chakrabarti07,eidelsberg2012} and
references therein.
In all cases, under the influence of ultraviolet radiation produced during the
stellar cycle, CO is the subject of a rich photochemistry, implying the
formation among other species, of its cation, CO$^+$.

CO$^+$ was the first molecular ion discovered outside the
Earth~\cite{fowler1909}, has been identified as a major constituent
of dense interstellar molecular clouds~\cite{roueff2010}. CO$^+$ also occurs in
flames~\cite{krupente65,seaton66,mitchell85}, and is expected to be very
abundant in the plasmas
formed by the hypersonic entry of spacecrafts and comets in the Martian
atmospheres~\cite{bultel2006,park1994}.
Cold plasmas containing a large number of molecules and molecular ions
constitute a subject of rising scientific interest, involving
more and more technological applications. CO$^{+}$ is of
the most important of such molecular ionic species, since
it occurs in practically all air assisted processes. Its dissociative
recombination (DR) can be regarded as the main photochemical carbon loss process
from the Martian atmosphere~\cite{fox1999} and it is considered the major
source 
of excited C$(^1D)$ atoms.

The kinetic description of all the above mentioned environments requires a good
knowledge of the rate
coefficients of the dominant reactions, including those between electrons and
molecular ions.
As for the CO$^{+}$ ions, their abundance is strongly affected by the DR,

\begin{equation}
 \mbox{CO}^{+}(v_{i}^{+}) + e^{-}\rightarrow  \mbox{C} + \mbox{O},
\label{eq:dr}
\end{equation}

\noindent
but also by other related competitive processes, such as the {\it inelastic}
($v_{f}^{+}>v_{i}^{+}$) and the {\it superelastic} ($v_{f}^{+}<v_{i}^{+}$) {\it
collisions} with electrons:

\begin{equation}
 \mbox{CO}^{+}(v_{i}^{+}) + e^{-}\rightarrow  \mbox{CO}^{+}(v_{f}^{+}) + e^{-}.
\label{eq:diff}
\end{equation}

\noindent
Here $v_{i}^{+}$ and $v_{f}^{+}$ stand for the initial
and final vibrational quantum number of the target ion, and rotational structure
was neglected.

The pioneering theoretical study of the DR rate coefficient made by 
\cite{guberman-dr07} shows that the dominant routes for the DR process are
curves of the $^3\Pi$, $^1\Pi$ and
$^1\Sigma^+$ symmetries. Guberman calculated the rate constants at $300$ K  to be
$4.2\times 10^{-7}$~cm$^3/$s for $^{12}$C$^{16}$O$^+$ and $2.9 \times 10^{-7}$
cm$^3/$s for
$^{13}$C$^{16}$O$^+$ respectively.
The most recent storage ring experiments of~\cite{rosen98}
gave a rate
coefficient of $2.75 \times 10^{-7}$~cm$^3/$s for $^{13}$C$^{16}$O$^+$ which is
in good
agreement with the calculations of Guberman. The earlier afterglow experiments
\cite{geoghegan91}
for $^{12}$C$^{16}$O$^+$  gave a rate coefficient of $1.6 \times
10^{-7}$~cm$^3/$s, which
does not agree with the theoretical calculations for $v^+_i=0$   and may indicate that a
proportion of
the CO$^+$ in that experiment was vibrationally excited.

Recently, a series of \textit{ab initio} calculations of the CO states
relevant to DR (and of the couplings between them) have been performed by Chakrabarti and Tennyson~\cite{chakrabarti06,chakrabarti07} using the $R$-matrix
 method. The
data obtained makes a new Multichannel Quantum Defect Theory (MQDT)
investigation of the process over the whole
energy range explored by the experiments \cite{rosen98,mitchell85} possible.

The main goal of the present work is to evaluate the DR cross sections and the
thermal rate using the previously mentioned {\it ab initio} data of~\cite{chakrabarti06,chakrabarti07}.
The paper is structured in the following way: Section~\ref{sec:theory} outlines
the main ideas and steps of our MQDT approach. Section~\ref{sec:COdata} presents
the molecular data used in the calculation. The main results are given in
section~\ref{sec:DRcs} and the paper ends by conclusions.

\section{The MQDT-type approach to DR}{\label{sec:theory}}

The MQDT approach~\cite{seaton83,gj85,jungen96,giusti80} has been shown to be a
powerful method for the evaluation of the cross
sections of the DR process. Although it was applied with a great success to
several diatomic systems like H$_2^+$ and its isotopologues
\cite{giusti83,ifs-a09,takagi93,tanabe95,ifs-a18,amitay99}, O$_2^+$
\cite{ggs91,guberman-dr99}, NO$^+$~\cite{sn90,ifs-a22,ifs-a36}
and triatomics like H$_{3}^{+}$ \cite{ifs-a26,kokoou01,kokoou03}, its
application to vibrational transitions,
mainly to superelastic collisions for NO$^+$ is relatively recent
\cite{Ngassam1,motapon06b}.

We aim to describe the sensitivity of the reactive electron-cation collisions on
the {\it vibrational} levels involved. At
least to a first approximation,
rotational effects are known to be negligible forCO$^+$ \cite{ifs-a22}.
The reasons for this rely on the weak Rydberg-valence interaction
responsible for the indirect process:
since the rotational structures and interactions play their role especially
within the indirect mechanism - due to its resonant character - the weakness of
this latter process
implies that the rotational effects - if of any relevance - can be roughly
restricted to the existence
of the centrifugal barrier, due to the rotational excitation. But even in this
latter context, rotation
does not matter very much, since the target ion and the neutral are
{\it equally} excited and, consequently, the Franck-Condon overlaps (direct
process) do not change significantly with respect to those occurring for the
case of rotationally ground states. However, this does not mean that electron-impact rotational
excitation is negligible~\cite{faure01}.

The theoretical summary given below is limited to an account of the
{\it vibrational} structure and coupling, illustrated mainly for DR.
However, the
reader should keep in mind that the other competitive reactions -- such as {\it
superelastic collision} (SEC) ($v^+_i>v^+_f$ in eq.~(\ref{eq:diff})), {\it
elastic collision} (EC) ($v^+_i=v^+_f$) and
{\it inelastic collision} (IC) ($v^+_i<v^+_f$) -- also occur and can  display quite similar features.

The DR can take place following two mechanisms:

\begin{enumerate}[label=(\roman*), labelindent=\parindent, leftmargin=*,
widest=iii]
 \item {the {\it direct} process where the capture takes place into a
dissociative
state of the neutral system (CO$^{**}$),

\begin{equation}
 \mbox{CO}^{+}(v_{i}^{+}) + e^{-}\rightarrow \mbox{CO}^{**} \rightarrow 
\mbox{C} + \mbox{O},
\label{eq:dra}
\end{equation}
}
 \item {the {\it indirect} process where the capture occurs \textit{via} a
Rydberg state of the molecule CO$^{*}$ which is  predissociated by the CO$^{**}$
state,

\begin{equation}
 \mbox{CO}^{+}(v_{i}^{+}) + e^{-}\rightarrow \mbox{CO}^{*} \rightarrow
\mbox{CO}^{**}  \rightarrow  \mbox{C} + \mbox{O}.
\label{eq:drb}
\end{equation}
}
\end{enumerate}
\noindent
 We note here, that we follow the standard nomenclature in this work, namely 
CO$^{**}$ and CO$^{*}$ represent the doubly excited and singly excited
 states of CO respectively. 
In both these processes autoionization is
in competition with the predissociation and leads, through the  reaction
(\ref{eq:diff}), to
 SEC, EC or IC. 

The MQDT treatment of DR involves ionization channels (describing the
electron-ion scattering)
and dissociation channels (describing the atom-atom scattering). Each ionization
channel
consists of a Rydberg series of excited states, extrapolated above the continuum
threshold
- a vibrational level $v^+$ of the molecular ion.  A channel is considered {\it
open} if its corresponding
threshold is situated {\it below} the total energy of the system, and {\it
closed} in the
opposite case. In the present work we have $N^+=N$, and as for the dissociative
channels, only open channels are used.

\noindent

The present MQDT approach is based on a description of molecular states in which
only
part of electronic Hamiltonian is diagonalized, within subspaces of
electronic states with similar nature. Moreover we use a quasidiabatic
representation
of molecular states \cite{sidis} to cope with problems due to the avoided
crossings of the potential energy curves. The short-range electronic
interactions between states of different subspaces are then represented in terms
of an electronic coupling operator \Vmat\ given by

\begin{equation}
\mathcal{V}_{d_{j},lv}^{\Lambda}(E',E)=\langle\chi_{{d}_{j}}^{\Lambda}
(R)|V_{{d}_{j},l}^{\Lambda}(R)|\chi_{v}^{\Lambda}(R)\rangle,
\label{coupling}
\end{equation}
\noindent
which couples the ionization channels (labeled by $lv$) to the dissociative
channels (labeled by $d_j$). Starting from \Vmat, one can build the short-range
reaction matrix \Kmat, which is a solution of the Lippmann-Schwinger
integro-differential equation:

\begin{equation}
\label{eq:Lippmann-Schwinger}
\boldsymbol{{\cal K}}=\boldsymbol{{\cal V}} + \boldsymbol{{\cal V}}
\frac{1}{E-\boldsymbol{H_0}} \boldsymbol{{\cal K}}.
\end{equation}
\noindent
{\boldmath $H_0$} being the zeroth order Hamiltonian associated to the molecular
system, i.e. the Hamiltonian operator excluding the interaction potential \Vmat.
The short-range effects are valid in the region of small electron-ion and
nuclei-nuclei distances, namely in the `A-region' ~\cite{ja77} where the
Born-Oppenheimer representation is appropriate for the description of the
colliding system. Here, the energy-dependence of the electronic couplings
can be neglected. In the case of weak coupling, a perturbative solution of
equation~(\ref{eq:Lippmann-Schwinger}) can be obtained.  This solution has been
proved to be exact to second order, in the case of energy-independent
electronic couplings \cite{Ngassam2}. 
In the external zone, the `B-region'~\cite{ja77}
represented by large electron-core distances, the Born-Oppenheimer model is no
longer valid for the ionization channels and a close-coupling representation in
terms of `molecular ion $+$ electron' is more appropriate. This corresponds to a
frame transformation defined by the projection coefficients:

\begin{equation}
\label{eq:coeffCv}
{\cal C}_{lv^{+}, \Lambda \alpha}=\sum_{v} U_{lv,\alpha}^{\Lambda}\langle
\chi_{v^{+}}| \cos(\pi \mu_{l}^{\Lambda}
(R)+\eta_{\alpha}^{\Lambda})|\chi_{v}^{\Lambda}\rangle,
\end{equation}
\begin{equation}
\label{eq:coeffCd}{\cal C}_{d_{j},\Lambda \alpha}=U_{d_{j}\alpha}^{\Lambda}\cos
\eta_{\alpha}^{\Lambda},
\end{equation}
\begin{equation}
\label{eq:coeffSv} {\cal S}_{lv^{+}, \Lambda
\alpha}=\sum_{v}U_{lv,\alpha}^{\Lambda} \langle \chi_{v^{+}}|
\sin(\pi\mu_{l}^{\Lambda}
(R)+\eta_{\alpha}^{\Lambda})|\chi_{v}^{\Lambda}\rangle,\end{equation}
\begin{equation}
\label{eq:coeffSd} {\cal S}_{d_{j},\Lambda\alpha}=U_{d_{j}\alpha}^{\Lambda}\sin
\eta_{\alpha}^{\Lambda}.
\end{equation}
\noindent
Here, $\chi_{v^+}$ is the vibrational wavefunction of the molecular ion,
and $\chi^{\Lambda}_{v}$ is a vibrational wavefunction adapted to the
interaction (A) region. The
index $\alpha$ denotes the eigenchannels built through the {\it diagonalization}
of the reaction matrix \Kmat\ in equation~(\ref{eq:Lippmann-Schwinger}) and
$U^\Lambda_{lv,\alpha}$ and $\eta_{\alpha}^{\Lambda}$ are related to the
corresponding
eigenvectors and eigenvalues, while $\Lambda$ refers to the electronic symmetry
of the
neutral species
($^1\Sigma^+$, $^1\Pi$, and $^3\Pi$ in the present study). 

The projection coefficients shown in
(\ref{eq:coeffCv})-(\ref{eq:coeffSd}) include the two types of couplings
controlling the process:
the {\it electronic} coupling, expressed by the elements of the
matrices \Umat\ and \etamat, and the {\it non-adiabatic} coupling between the
ionization channels, expressed by the matrix elements involving the quantum
defect
$\mu_{l}^{\Lambda}$. This latter interaction is favored by the variation of the
quantum defect with internuclear distance $R$. The matrices \Cmat\  and 
\Smat\
with the elements given by (\ref{eq:coeffCv}) to (\ref{eq:coeffSd}) are the
building blocks of
the `generalized' scattering matrix \Xmat:
\begin{equation}
\label{eq:Xmatrix}\Xmat =\frac{\Cmat+i\Smat}{\Cmat-i\Smat},
\end{equation}
whereas the `proper' scattering matrix, restricted to the {\it open} channels,
is given by~\cite{seaton83}:
\begin{equation}
\label{eq:elimination}\SSmat=\Xmat_{oo}-\Xmat_{oc}\frac{1}{\Xmat_{cc}-\exp({\rm
-i 2 \pi} \numat)} \Xmat_{co}.
\end{equation}
\noindent
More precisely the physical S-matrix is obtained from the $2\times2$
sub-matrices of \Xmat\  involving 
the open (o) and closed (c) channels, and from the diagonal matrix \numat\ 
containing the effective quantum numbers ${\nu}_{v^{+}}=[2(E_{v^{+}}-E)]^{-1/2}$
(in atomic units) associated with each vibrational threshold $E_{v^{+}}$ of the
ion situated {\it above} the current energy $E$. 
For a molecular ion initially in  vibrational level $v_{i}^{+}$ and
recombining with an electron of energy $\varepsilon$, the cross section for
capture into {\it all} the dissociative states $d_{j}$ of the same symmetry
${\Lambda}$ is given by:
\begin{equation}
\label{eq:cs-partial}
\sigma^{{\Lambda}}_{{\rm diss} \leftarrow v^{+}_{i}}=\frac {\pi
{\rho}^{\Lambda}}{4\varepsilon}\sum _{j} \sum _l |S^{{\Lambda}}_{d_{j}
\leftarrow l v^{+}_{i}}|^{2},
\end{equation}
\noindent
where $\rho^{\Lambda}$ is the ratio between the multiplicities of the neutral
system and of the ion.
One has to perform the MQDT calculation for each group of dissociative states of
symmetry $\Lambda$, and the sum over the resulting cross sections is the total
DR cross section:
\begin{equation}
\label{eq:cs-total}\sigma_{{\rm diss} \leftarrow
v^{+}_{i}}=\sum_{\Lambda}\sigma^{\Lambda}_{{\rm diss} \leftarrow v^{+}_{i}}
\end{equation}
In a similar way, the cross section for a vibrational transition of a molecular
ion from  the initial level $v_{i}^{+}$ to the final level $v_{f}^+$ is
(excitation and/or deexcitation):

\begin{equation}
\label{eq:csvib-partial} 
\sigma^{{\Lambda}}_{{ v_{f}^{+}} \leftarrow v^{+}_{i}}= \frac {\pi}{4} \frac
{{\rho}^{{\Lambda}}}{\varepsilon} \sum _{l,l'} |S^{{\Lambda}}_{l'
v_{f}^{+}\leftarrow l v^{+}_{i}}|^{2},
\end{equation}
\noindent
while the total cross section for this vibrational transition reads as:
\begin{equation}
\label{eq:csvib-total} 
\sigma_{{ v_{f}^{+}} \leftarrow v^{+}_{i}}=\sum_{\Lambda} \sigma^{\Lambda}_{{
v_{f}^{+}} \leftarrow v^{+}_{i}}.
\end{equation}
\noindent

\section{Molecular data}{\label{sec:COdata}}
\subsection{\textit{Ab initio R-matrix calculations}}{\label{subsec:rmatrix}}

The earliest attempts to compute excited states of CO relevant for the
dissociative recombination and their autoionization widths goes back to 1996
\cite{tennyson1996a,tennyson1996b} but were restricted to a single geometry.

Ten years later,  two of us~\cite{chakrabarti06,chakrabarti07} performed a much
more detailed R-matrix
calculation for a series of fixed geometries. These calculations
used up to the $14$ lowest CO$^+$ states in a  close coupling expansion
where each of these states were represented using a valence complete
active space configuration interaction expansion. The calculations
were repeated for $10$ different internuclear distances in the range of $1.5~
a_0$ to $3.5
~a_0$.
Moreover \cite{chakrabarti07} computed positions and widths for
a number of dissociative states, which appear as
resonances in the R-matrix scattering calculations, as a function
of geometry. In particular, resonance states of electronic symmetry
$^{1}\Sigma^+$, $^{1}\Pi$ and $^{3}\Pi$ that have been used
in the present work as the other symmetries are only very weakly coupled.

Although the R-matrix resonance calculations \cite{chakrabarti07} for the CO
molecule
have not been subject to comparisons and calibrations against all
available spectroscopic data, some Rydberg states previously were calculated~\cite{tennyson1996a,chakrabarti06}. 
It was found that these Rydberg states are placed higher than the observed
values which is consistent with the variational nature of the R-matrix
calculations.  As a consequence, for low $\ell$ values, the estimated quantum
defects were uniformly
lower by about $0.07$ compared to the available experimental data, see for
example Tables 4 and 5 in \cite{chakrabarti06}.
The geometry dependent study on CO also showed that the quantum
defects depend weakly on geometry except when the adiabatic curves are
perturbed by an avoided crossing. This approach was successfully used by \cite{ifs-a36} to give
greatly improved results for DR of NO$^+$.

\subsection{\textit{Modeling of the dissociative curves}}\label{subsec:disspec}

\begin{figure}[t]
\centering
\includegraphics[width=0.95\columnwidth]{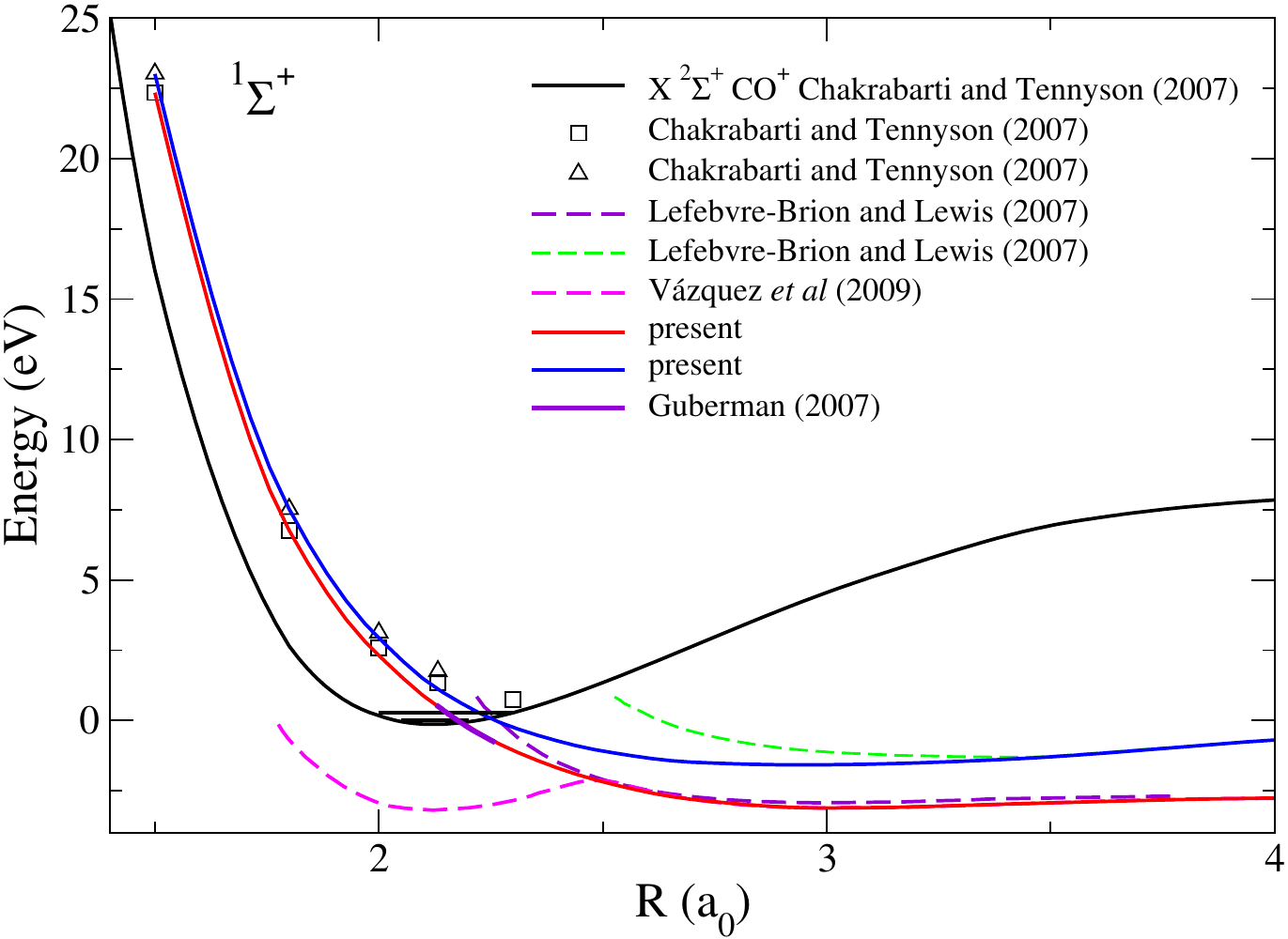}
\caption{Electronic states relevant for dissociative recombination with
$^1  \Sigma^+$  total symmetry. The ion ground state
is shown in black; $\square, \vartriangle$ are the R-matrix data~\cite{chakrabarti07}: (Chakrabarti and Tennyson 2007) for the lowest two CO$^{**}$ states. The two continuous colored curves crossing the ion are the
dissociative states obtained as outlined in section~\ref{subsec:disspec}. The
curves indicated in the
legends refer to the curves from~\cite{guberman-dr07}: Guberman (2007), \cite{lefebvre07}: Lefebvre Brion and Lewis (2007) and \cite{Vazquez}: Vazquez {\it et al} (2009)}
\label{pecsigma}
\end{figure}

The DR cross section is extremely sensitive to the position of the potential
energy curves (PECs) of the neutral dissociative states with respect to that of
the target ion. More specifically, a slight change of the crossing point of the
PEC of a neutral dissociative state with that of the ion ground state can lead
to a significant change in the predicted DR cross section. The magnitude of this
change is roughly proportional to the
square of the Franck-Condon type integral appearing in equation
(\ref{coupling}).
This is an important issue here as we
use the approximate PECs of the
dissociative states above the ion ground state produced by R-matrix
calculations \cite{chakrabarti06,chakrabarti07}. Indeed, for example,
the crossing point of the lowest relevant $^1\Sigma^+$ state was found to be at
 higher energy and at larger internuclear distance than those obtained 
previously using a de-perturbation procedure \cite{tchang92} or {\it ab-initio} quantum
chemistry computations \cite{guberman-dr07}. Moreover, recent calculations
~\cite{Vazquez,lefebvre07,lefebvre10}  give PECs for the
\emph{lowest} neutral states relevant for DR,
that only partially agree with those of the other calculations mentioned
above. 

All the available lower dissociative neutral states for the $^1\Sigma^+$
symmetry is shown in Figure~\ref{pecsigma}, 
where we have presented
the PECs relevant for the present study. The zero of energy was chosen to be
the lowest vibrational level of the ground electronic state of the molecular
ion.
At the same time, apart from the dissociative curves, the MQDT calculation also
requires the couplings between the ion ground state and the neutral
dissociative states. The R-matrix calculations of \cite{chakrabarti07} 
systematically provide autoionization widths
for a large number of
states which contain information about the required couplings. Such
comprehensive data on couplings are not available from any other source. 
Thus  use their data in our dynamical study on DR but 
correct the PECs when it is needed. In
this way we undertake a semi-empirical calculation.

The dissociative (valence) states produce numerous avoided crossings when
crossing the series of Rydberg states. 
By turning to the diabatic representation these avoided crossings will become
"true" crossings 
essential for the {\it indirect} process. 
A quasi-diabatic PEC may be constructed, by
smoothly connecting the resonant states obtained by scattering calculations to
the asymptotic adiabatic states. A more detailed inspection of the available
material, both diabatic and adiabatic, confirms the quality of the diabatic
states obtained by \cite{guberman-dr07}, justifying their use in the
construction of
quasi-diabatic PECs by matching them in the relevant range of internuclear
distances. As a general rule, we apply a slight shift to the R-matrix PECs
only if it is necessary in order to match them with the
diabatic curves of \cite{guberman-dr07}. This is clearly the case for
the
first state of $^1\Sigma^+$ symmetry, see Fig~\ref{pecsigma}. No such a behavior
 is visible for the second dissociative PEC, but we apply the same shift
in quantum defect as we
applied to connect the first state  to its quasi-diabatic counterpart.
Correlation to the proper asymptotic limits follows both the \cite{wigner28} rules
and the adiabatic curves of \cite{Vazquez}, where they are available.
In our model, we
construct the lowest dissociative states as a function of the internuclear
distance on the basis of a
physically reasonable compromise between the available valence curves coming
from all
the above cited authors.
As an example, consider 
the lowest state of $^{1}\Sigma^{+}$ symmetry:
At the smaller internuclear distances, our quasi-diabatic PEC follows the
R-matrix points of~\cite{chakrabarti07}.
At larger internuclear distances we shift progressively the R-matrix
dissociative curve  in order to match them with those of
\cite{guberman-dr07}. This curve is used until its crosses  the
PEC of the molecular ion,
and beyond, followed by a smooth connection to the lowest \emph{diabatic} PEC of
\cite{lefebvre07}, which agrees very well with
the asymptotic behavior of the lowest \emph{adiabatic} potential energy curves
of Vazquez at large internuclear distances.
The dissociative
curves constructed in this way will agrees well with the PEC used by Guberman. 
A similar procedure is followed for all the other excited states with different
symmetries.

\begin{figure}[t]
\begin{center}
\includegraphics[width=0.95\columnwidth]{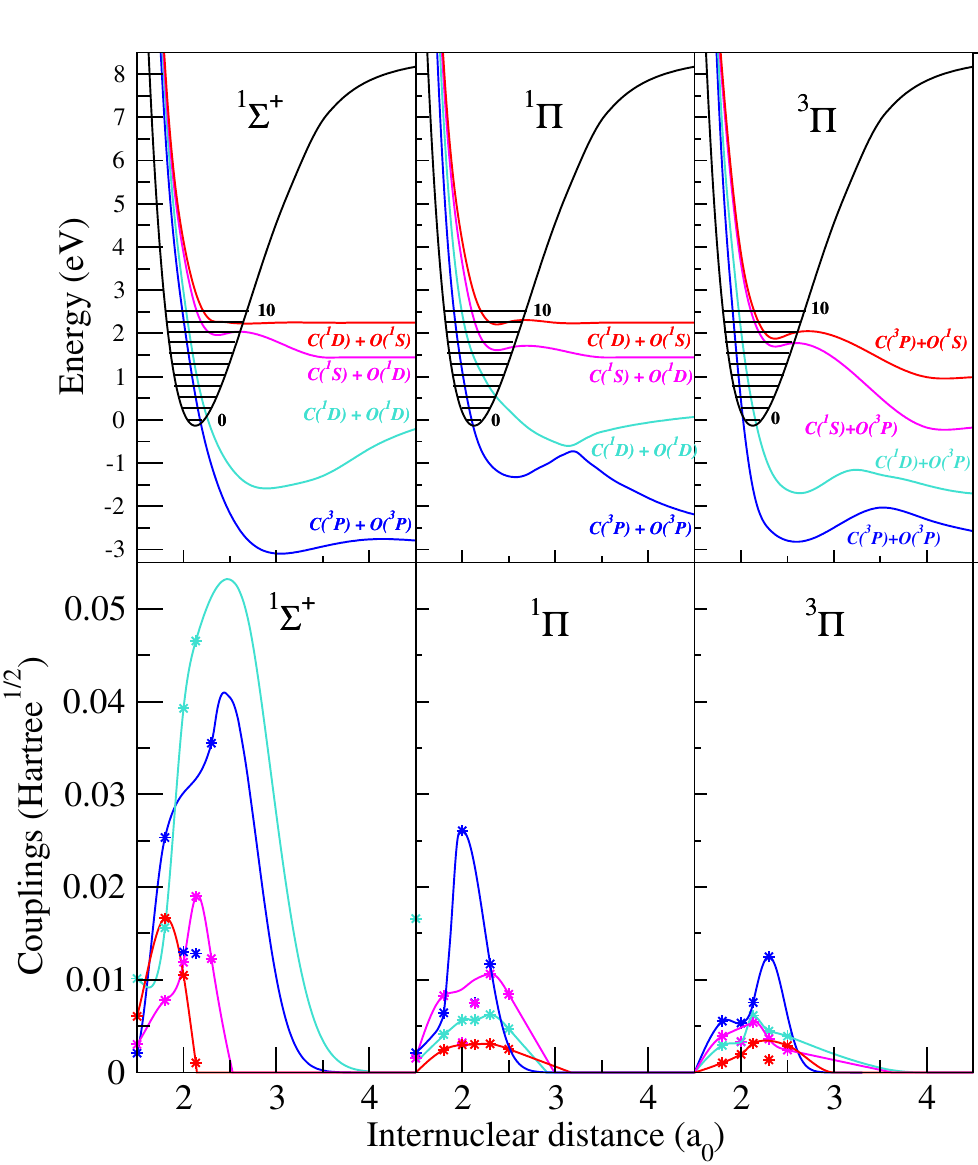}
\caption{The dissociative curves and couplings relevant for CO$^+$
dissociative recombination for the symmetries indicated in each figure. Top
panel: the adjusted potential energy curves (cf section~\ref{sec:COdata}), the
ion ground state
and first 10 vibrational levels are represented in black. Bottom panel:
couplings between the valence dissociative states and the ionization
continua.}\label{curves}
\end{center}
\end{figure}

As for the highest R-matrix dissociative PECs, we believe that they are
robust, since the inherent errors in the R-matrix calculations  decrease
significantly within the degree of excitation.
Finally, Figure~\ref{curves} shows the ion ground state and vibrational levels,
the adjusted
dissociative curves used in the present calculations and their corresponding
autoionization widths coming from the R-matrix computations.

\section{Evaluation of the cross section using the MQDT-type
approach}{\label{sec:DRcs}}

\subsection{Mechanisms and couplings}\label{subsec:mech}

Using the set of molecular data (PEC and electric couplings) determined as
described in the previous section, we performed a
series of MQDT calculations of the DR cross section, assuming the molecular
ion to be initially in its electronic ground state ($X^{2}\Sigma^{+},
v_{i}^{+}=0,1,2$) and by neglecting rotational and spin-orbit effects. The
calculations were performed for
the states with total symmetry of $^{1}\Sigma^+$, $^{1}\Pi$ and $^{3}\Pi$,
while 
the number of dissociative states considered for each symmetry are indicated in
Figure~\ref{curves}.

We consider incident electron energies
from $0.01$ meV up to $3$ eV. Since the dissociation energy
of CO$^{+}$($X^{2}\Sigma^{+}$) is about $8.5$ eV, the majority of the 
$53$ vibrational
levels of the ion lie \textit{above} the total energy
of the CO$^+ + e$ system. These levels are associated with \textit{closed}
ionization channels, as defined in section~\ref{sec:theory}, responsible for
temporary
resonant capture into Rydberg states. As the energy increases, more and more
ionization channels become \textit{open}, which can result in autoionization
leading to competitive processes like IEC or SEC, decreasing the flux of DR.

The direct electronic couplings between ionization and dissociation channels -
equation (\ref{coupling}) - have been extracted from the autoionization widths
of
the valence states~\cite{chakrabarti07}. 

Since a highly accurate solution of the Lippman-Schwinger
system of integral equation (\ref{eq:Lippmann-Schwinger}) is difficult to
obtain~\cite{takagi2000,pichl2000}, we take advantage of the fact
that the
couplings (among ionization and dissociation channels) involved are small and
perturbative solution is acceptable. Following the main ideas of the
earlier studies~\cite{ggs91,ifs-a09}, we
adopt a second-order perturbative expansion, which accounts for all the
basic mechanisms involved in DR, including \textit{indirect} electronic
interaction between the ionization channels.

The non-adiabatic couplings between the ionization channels rely, see Eqs.
(\ref{eq:coeffCv}) and (\ref{eq:coeffCd}), on the $R$ dependence of the quantum
defects, which have been evaluated using the \textit{ab initio} calculations
described
in Section~\ref{subsec:rmatrix}. 

For each dissociative channel available, we have
considered its interaction with the most relevant series of Rydberg states,
namely for $^1\Sigma^+$ symmetry the lowest $4$ partial waves ({\it s, p, d} and
{\it f}) were considered, while for the $^{1}\Pi$ and $^{3}\Pi$ symmetries only
the lowest $3$ ({\it s, p, d}) have been used. For the sake of simplicity,
figure~\ref{curves} shows the so-called global couplings which were
constructed using the autoionization widths of~\cite{chakrabarti06,chakrabarti07}, namely 
$V^\Lambda_{d_{j}}=\left(\left(\sum_l\Gamma^\Lambda_{d_j,l}\right)/2\pi\right)^{
1/2}$.

According to Eq.~(\ref{eq:elimination}), the interference
between the \textit{direct} process (involving open channels exclusively and
described by the first term: $\boldsymbol{X_{oo}}$) and the \textit{indirect}
one (involving closed as well as open
channels and accounted for by the second term) results in what  the {\it
total}
process. 

\begin{figure}[t]
\centering
\includegraphics[width=0.95\columnwidth]{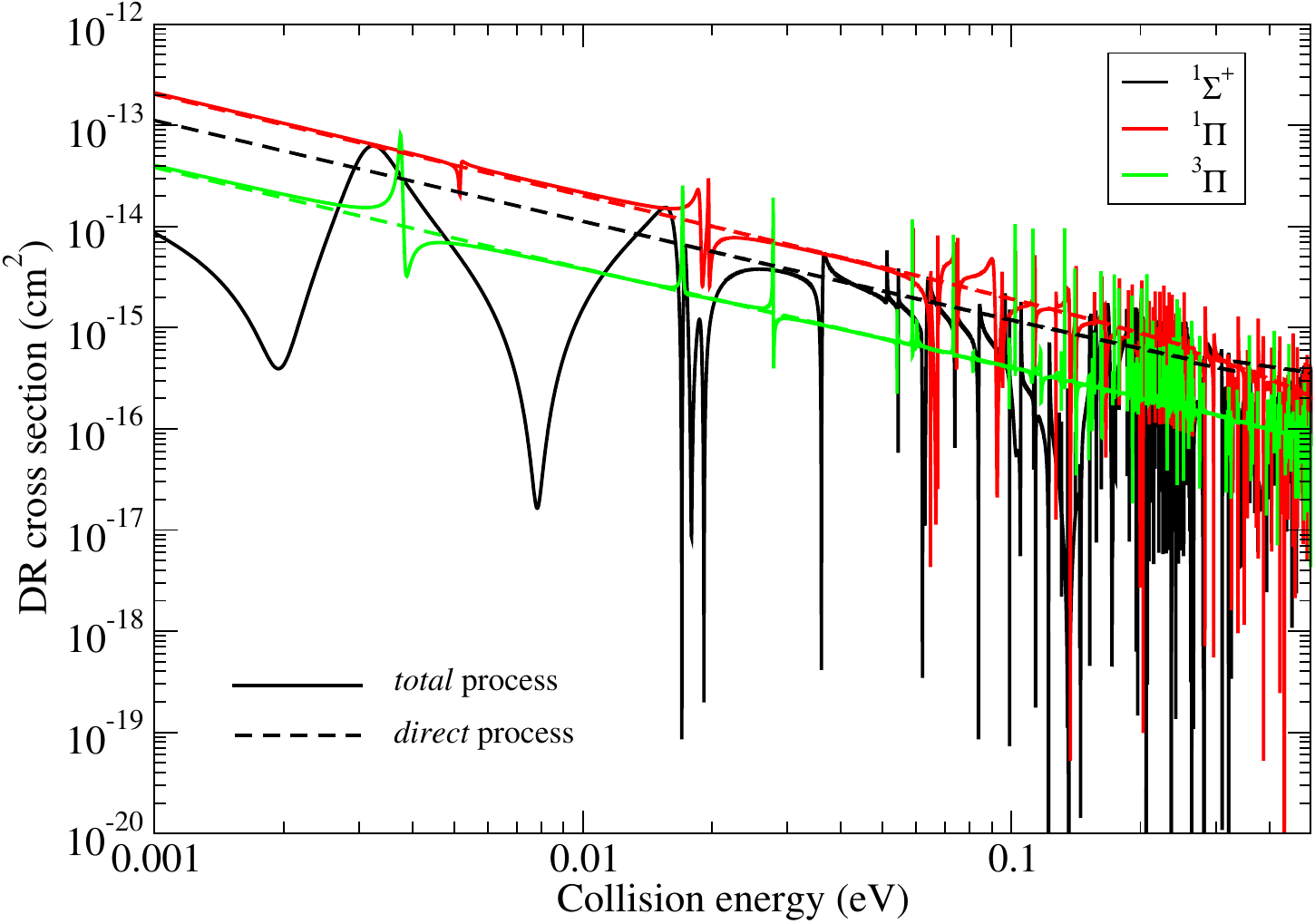}
\caption{Cross sections for the direct (dashed lines) and the total (direct $+$
indirect) processes (continuous lines), for a vibrationally relaxed target in electronic states with $^1\Sigma^+$
(black), $^1\Pi$ (red) and $^3\Pi$ (blue) symmetries.
}
\label{directglobal}
\end{figure}

The role of the direct and indirect mechanisms in the DR process can be seen in
figure~\ref{directglobal}. 
The full colored lines show the total cross sections for the different
symmetries used in the calculations, while dashed lines of the same color
represent the cross sections given by the direct process only. As one could
expect from the magnitude of the valence-Rydberg couplings, see
figure~\ref{curves}, a strong dependence on the states and interactions can be
observed, the majority  of the cross section is given by the $^1\Pi$ and
$^1\Sigma^+$ symmetries, while the $^3\Pi^+$ symmetry contributes only a small part. 
Moreover, for both states with $\Pi$ symmetry, the indirect process plays a
minor role for this ion; the magnitude of the total cross section is given by
the direct process and the indirect one is responsible for the resonance
structures. This is most striking for the $^1\Pi$ symmetry, where the total and
direct only cross sections lie on the top of each other. The importance of the
indirect mechanism is relevant for the $^1\Sigma$ symmetry, where due to the {\it
destructive} interference between the direct and indirect processes the total
cross section is lowered by almost one order of magnitude compared to the direct
one, at certain collision energies. The indirect processes give rise to broad
resonance structures in the range of the collision energy of the present study.
Even though the indirect process
globally plays a minor role, evaluating its effect gives increased insight into the
recombination mechanism.

The importance of the different dissociation paths in our calculation is
summarized in table~\ref{table}. Our calculated final state distribution are
compared with the experimental values of \cite{rosen98} for
four relative collision (detuning) energies. For the zero collision energy the
statistical error was evaluated to be about $5\%$, while for higher energy
values, due to the poor statistics, it is around $30\%$. 

\begin{figure}[t]
\centering
\includegraphics[width=0.95\columnwidth]{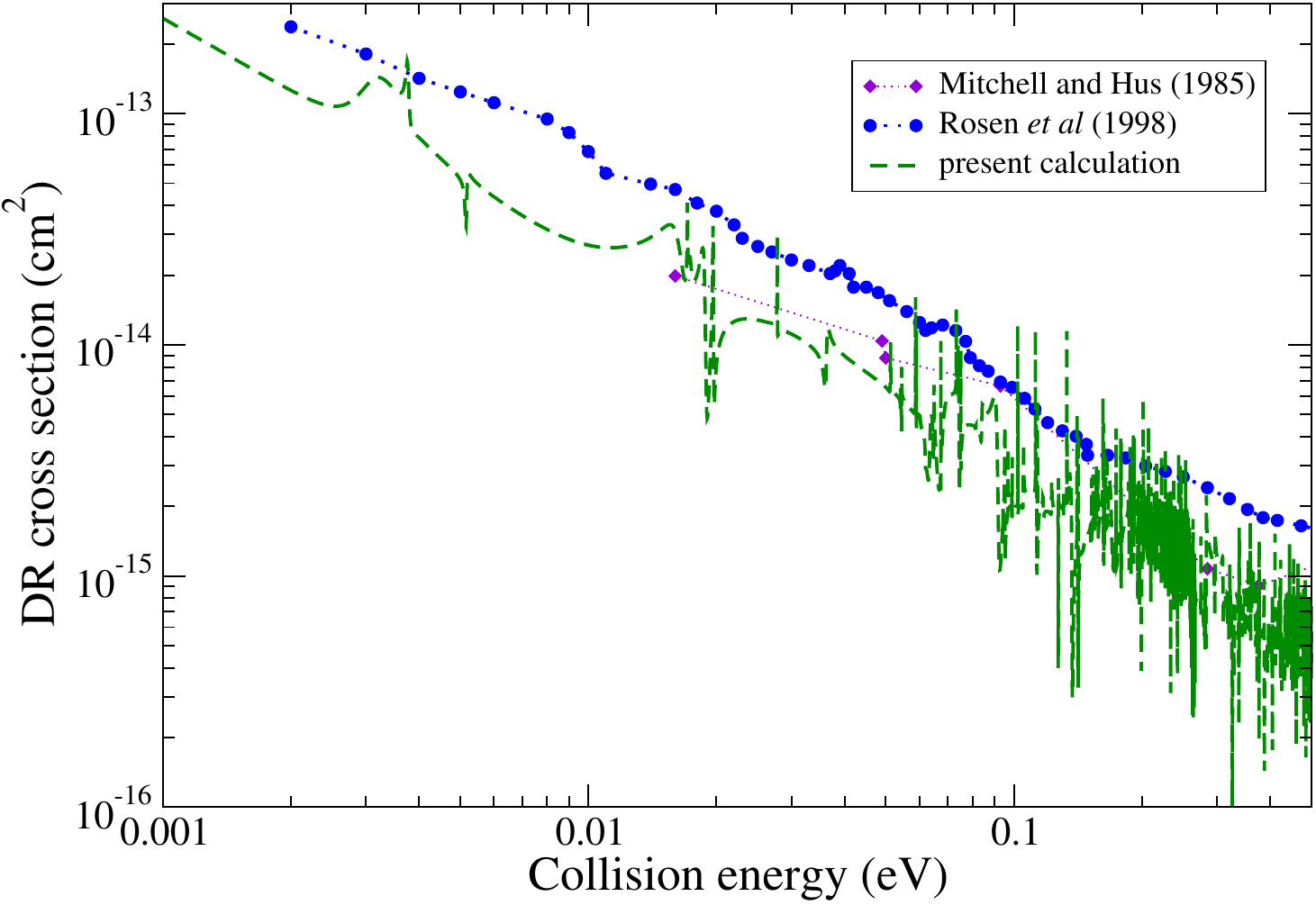}
\caption{Cross sections for the total (direct $+$ indirect) DR process for a vibrationally relaxed target including
the summed-up contributions of
the states with $^1\Sigma^+$, $^1\Pi$ and $^3\Pi$ symmetry. The experimental
results of \cite{mitchell85} are shown with a dotted line with
full violet diamonds, while those of \cite{rosen98} with dotted line with full blue circles, 
respectively.}
\label{total}
\end{figure}

\begin{table*}
\begin{center}
\caption{\label{table}The calculated final state distribution and its comparison
with the experimental values for four relative collision (detuning) energies.}
\begin{tabular}{ccccccccc}
\hline
Dissociation&\multicolumn{8}{c} {energy (eV)}\\
path & \multicolumn{2}{c}{10$^{-5}$} & \multicolumn{2}{c}{0.4} &
\multicolumn{2}{c}{1} & \multicolumn{2}{c}{1.5}\\
& calc & exp$^\dagger$ & calc & exp & calc & exp & calc & exp\\ 
\hline
C($^3$P)+O($^3$P) & 86.56\% & 76.1\% & 80.85\% & 53\% & 43.25\% & 39\% & 50.93\%
& 38\%\\
C($^3$P)+O($^1$D) & - & 9.4\% & - & 8\% & - &15\% &- & 11\%\\
C($^1$D)+O($^3$P)& 13.41\% & 14.5\% & 19.15\% & 34\% & 9.61\% & 35\% & 4.50\% &
35\%\\
C($^1$S)+O($^3$P)& & 0.0\% & & & <$10^{-5}$\% & 5\% & <10$^{-2}$\% & 5\%\\
C($^1$D)+O($^1$D)& & & & 5\% & 47.14\% & 6\% & 44.56\% & 11\%\\
C($^3$P)+O($^1$S)& & & & & & &$\sim$10$^{-11}$\%& \\
C($^1$S)+O($^1$D)& & - & & - & &-& $\sim$10$^{-7}$\%& -\\
\hline
$^\dagger$~\cite{rosen98}
\end{tabular}
\end{center}
\end{table*}

The dominant dissociation pathway for the DR of the CO$^{+}$ is the
one which correlates with the C($^3$P)+O($^3$P) atomic limits. Table~\ref{table}
shows a qualitative agreement between the calculated and the measured final
state distributions. The agreement is good, particularly at low collision energies.
At higher energies, the two sets of data disagrees, mainly due to the increased
experimental statistical errors and the fact that our 
model lacks a dissociating pathway correlating to the C($^3$P)+O($^1$D) atomic
limits. 

\section{Results and discussion}

\begin{figure}[t]
\centering
\includegraphics[width=0.95\columnwidth]{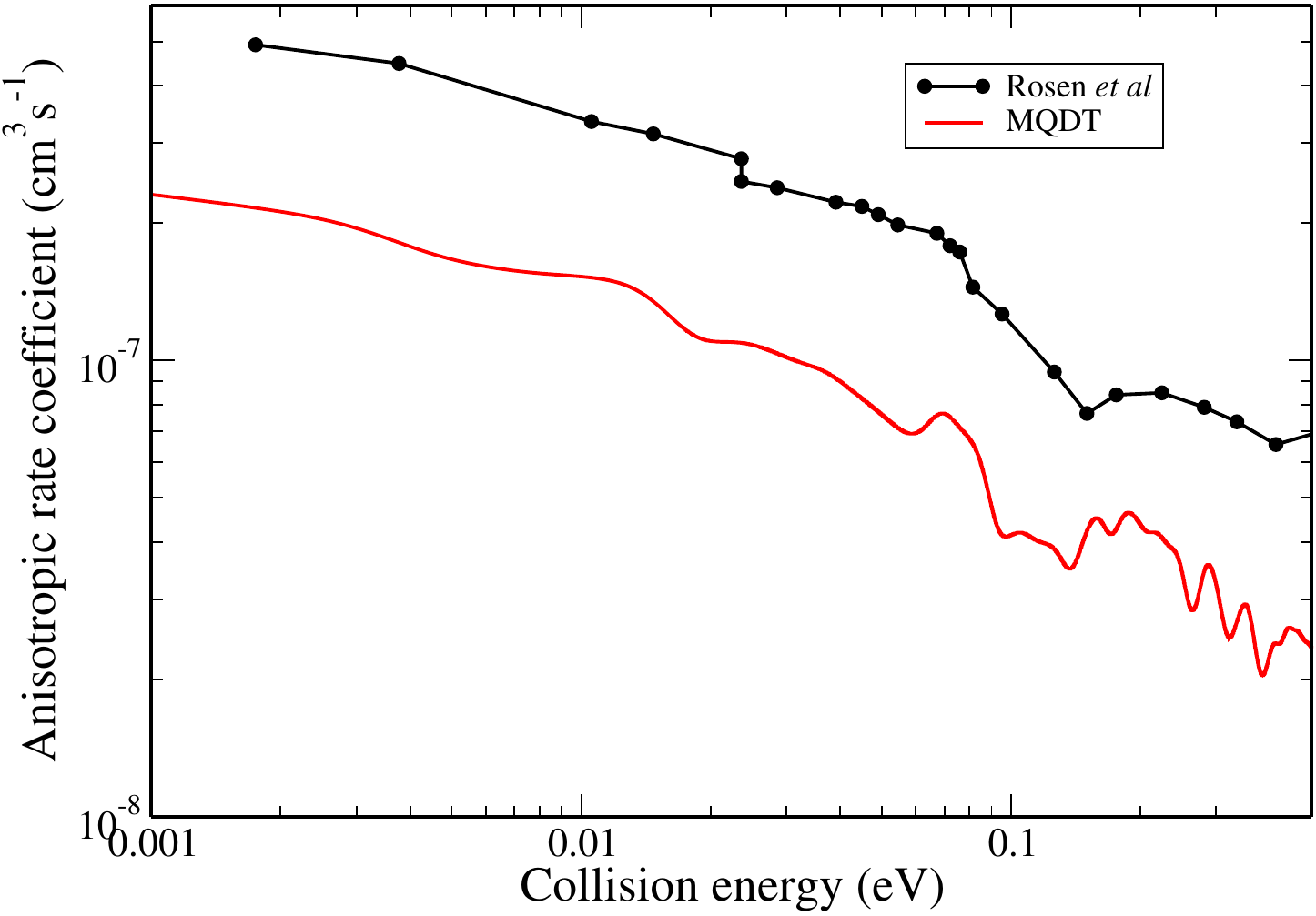}
\caption{Anisotropic rate coefficients for a vibrationally relaxed target. Experimental results~\cite{rosen98}
are in given in black with filled circles. Our theoretical results, for all
symmetries considered here, are shown in
red.
}
\label{conv_ani}
\end{figure}
\begin{figure}
\centering
\includegraphics[width=0.95\columnwidth]{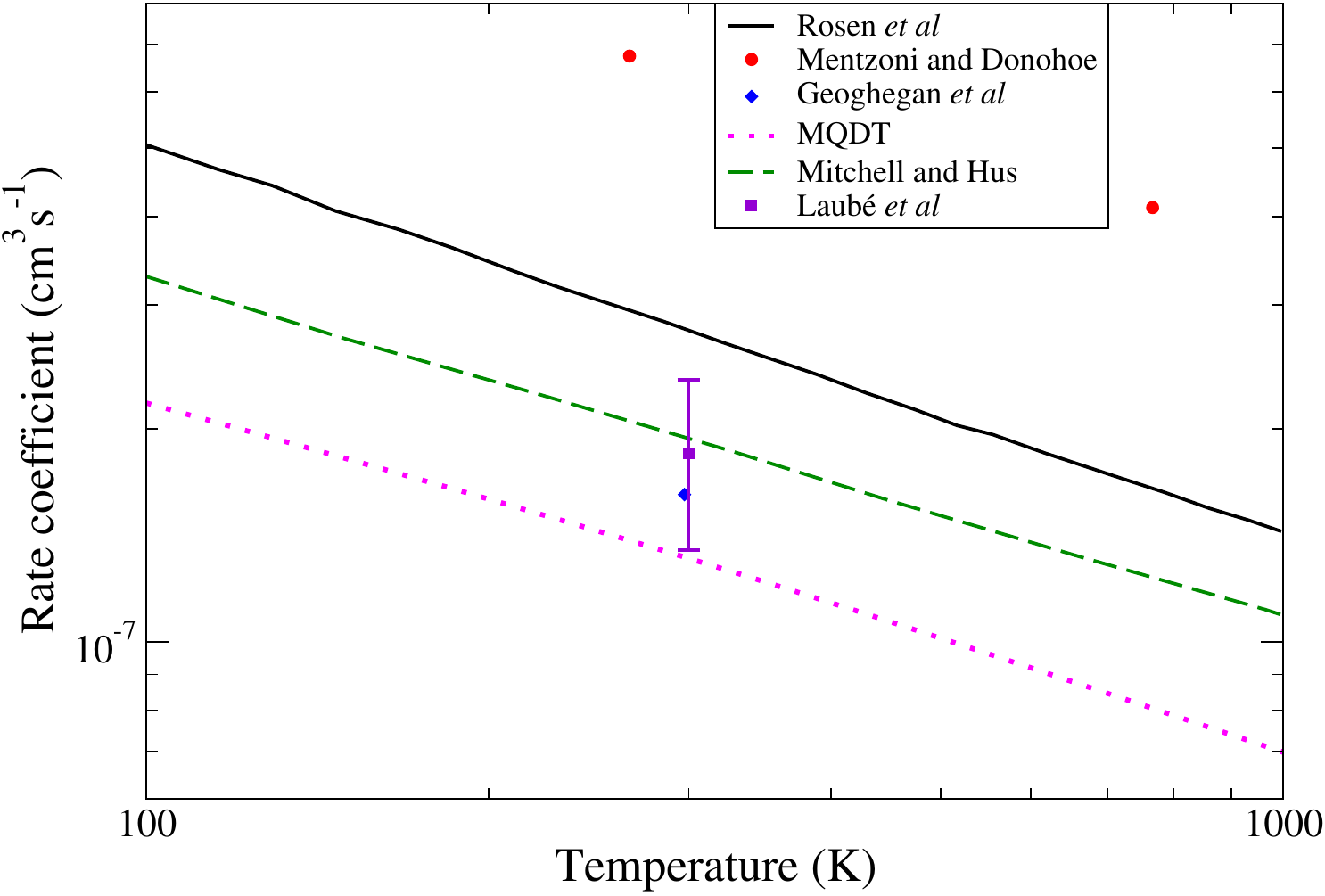}
\caption{Isotropic rate coefficients for  a vibrationally relaxed target. Our theoretical results, for all
symmetries considered
here, are represented by magenta dotted line. The experimental results are shown
as follows: \cite{mitchell85}, green dashed line; \cite{rosen98}, black solid line;
\cite{geoghegan91}, blue diamonds,  \cite{mentzoni68}, red circles and \cite{laube98}, violet squares.}
\label{conv_iso}
\end{figure}

The DR cross section corresponding to the three symmetries contributing to the
process ($^1\Sigma^+$, $^1\Pi$ and $^3\Pi$), summed according to Eq.~(\ref{eq:cs-total}),
are shown in figure~\ref{total}. The calculations have been performed for the $^{12}$C$^{16}$O$^+$ isotopologue of the cation.
The total cross section is
characterized by resonance structures, superimposed on a smooth background ({\it
direct} process only). The strong interaction between the closed channels,
associated with higher vibrational ion levels and the current dissociative
states, contaminates, via
non-adiabatic or indirect electronic coupling, the direct interaction between
the entrance and the dissociative channels. This results in a stronger capture
probability, and consequently a higher DR cross section, which leads to the
rich resonance structures observed in the total cross section. The direct DR cross
section is proportional to the square of the Franck-Condon overlap, of the
type seen in Eq.~(\ref{coupling}), between the vibrational wave functions of
the ion and the dissociation state of the neutral. The favorable crossings at
lower collision energies of the dissociation states correlating to the
C($^3$P)+O($^3$P) and C($^1$D)+O($^1$D) atomic limits (blue and cyan curves in
Fig.~\ref{curves}) with the ionic ground state (black curve on the same
figure) and the significant couplings to the entrance channel are responsible
for the large contribution by the direct DR cross sections to all the three
symmetries used in the calculations. This is the smooth background cross section
on which the contribution from the indirect process is superimposed, resulting
in the total cross section shown in Fig.~\ref{total}. The closed ionization channels
are those responsible for the temporary capture  into the Rydberg states and
consequently to the indirect mechanism. There is an increasing number of
resonance states of all symmetries at higher energies, and some  or all of these can
participate in the DR process via channel mixing mechanism.

The role of each dissociation  channel depends on the energy of the incident
electron. At higher collision energies, when the other two dissociation channels
become open (magenta and red curves in figure~\ref{curves}) due to their
less-favorable crossings with the ionic ground state the indirect process gains
importance leading to the resonance structures in the total cross section. We
have included all the contributions of the dissociative channels in our
calculations. 

In view of the comparison with the storage-ring measurements, we have convoluted
our MQDT cross
sections with the {\it anisotropic} Maxwell velocity distribution function for
the electrons. The transversal and longitudinal
temperatures  are $K_{B}T_{\perp}=1$ meV and $K_{B}T_{\parallel}=10$ meV,
respectively. The calculated anisotropic rate coefficient (magenta) together
with the measured one (green) \cite{rosen98} are shown in figure \ref{conv_ani}.
The overall agreement with experimental value is good, our calculated rate
coefficient reproduces perfectly the trends of the experimental curves,
underestimating them on average by a factor of $2$. 

\begin{figure}[t]
\centering
\includegraphics[width=0.95\columnwidth]{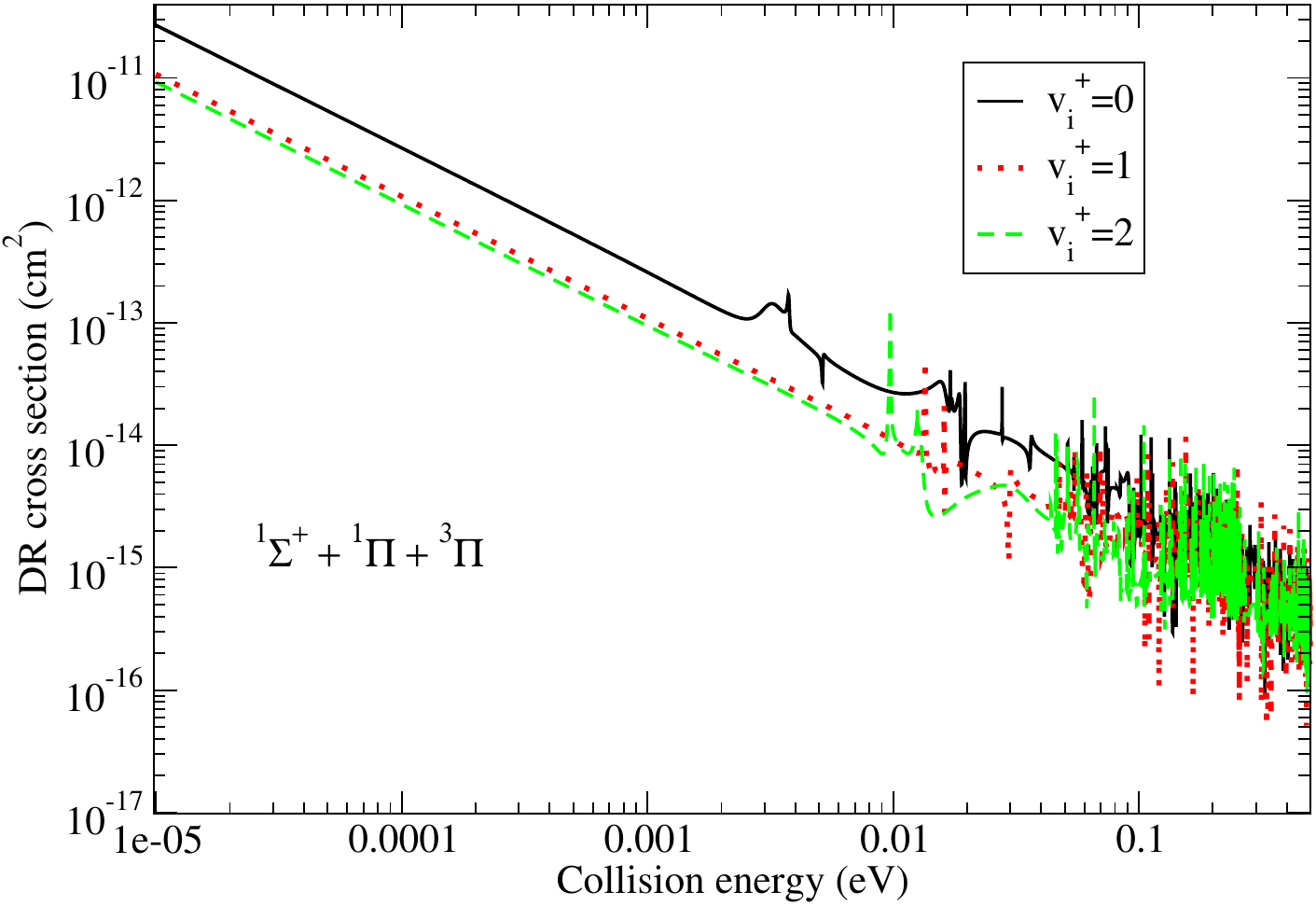}
\caption{Total DR cross sections for different initial vibrational levels of the
ion, for all the considered symmetries. Black continuous line stand for
$v_i^+=0$, while the red dotted and green dashed lines for  $v_i^+=1$ and $2$,
respectively.
}
\label{dr_cs_vi}
\end{figure}

\begin{figure}[t]
\centering
\includegraphics[width=0.95\columnwidth]{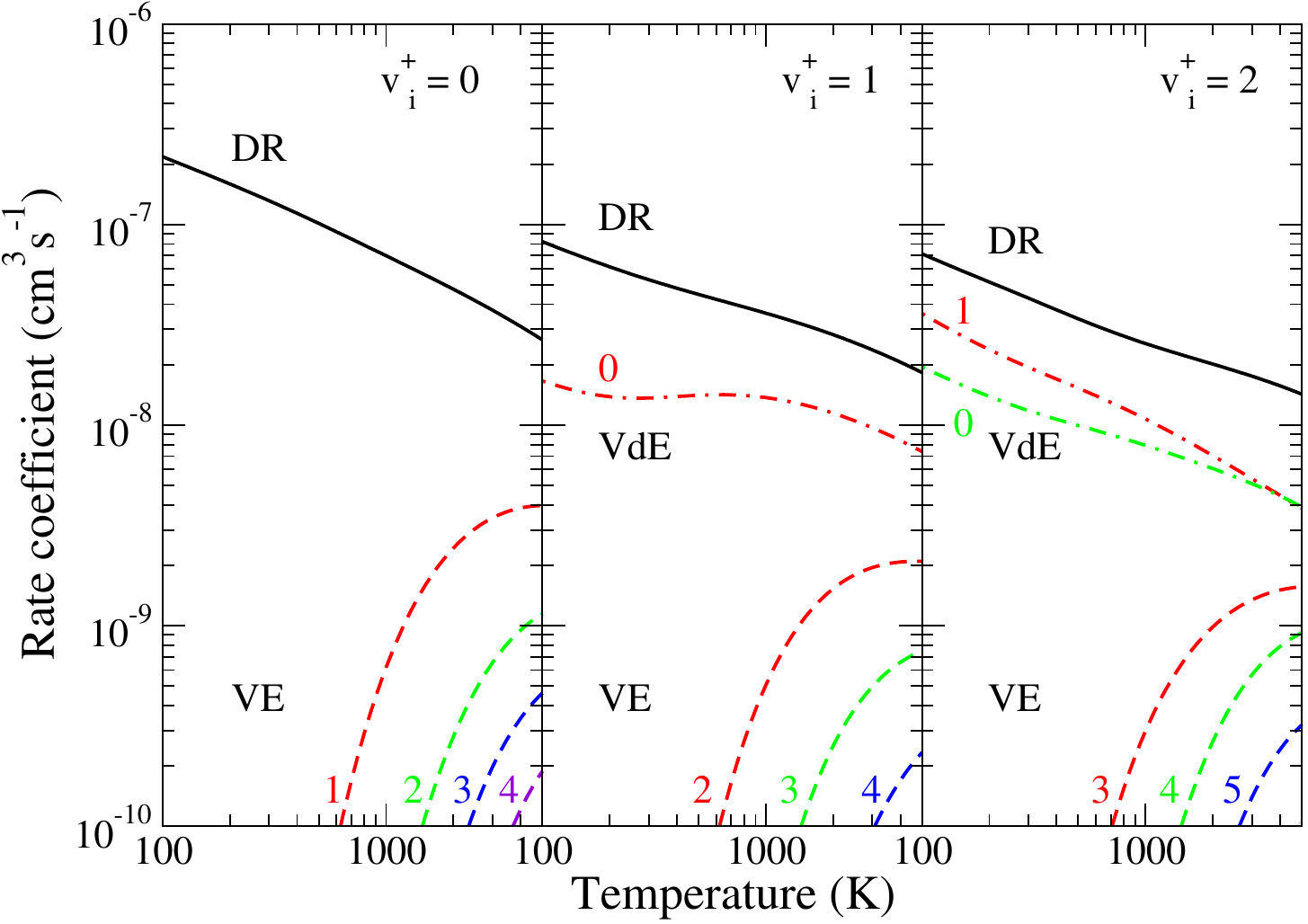}
\caption{
DR and state-to-state VE and VdE isotropic rate coefficients of CO$^+$ in its
ground electronic state, $v^+_i$ standing for the vibrational quantum number of
the target ion. Curves of same color show the rate coefficients for the
vibrational (de-)excitations corresponding to the same $|\Delta v| = |v^+_f -
v^+_i|$, $v_f > v_i$ for the VE and $v_f < v_i$ for the VdE global rate
coefficients, respectively. The final vibrational quantum numbers of the ion are
indicated for these processes.
}
\label{rate_vi}
\end{figure}

In addition to the anisotropic rate coefficients, starting from the computed
cross section, we have evaluated the Maxwell \textit{isotropic} rate coefficient
as well, for a broad range
of electronic temperatures, relevant especially for cold non-equilibrium
plasmas. The rate for $v_i^+=0$ is displayed in figure \ref{conv_iso}, and its
thermal ($300$ K) value is in excellent
agreement with the measured values obtained in merged-beam~\cite{mitchell85} and \cite{laube98} as
well as storage-ring based~\cite{rosen98} collision experiments. The rate coefficients of \cite{mentzoni68} presented as red circles in the figure are placed higher than the other measurements, probably due to the presence of CO$^+\cdot ($CO$)_n$ clusters in the sample, as it was pointed out by \cite{whitaker81}. 
Moreover, our calculations have shown much less isotopic effect than those found by \cite{guberman-dr07}, having as mean deviation $0.43\%$ (due to the different nuclear reduced masses)  among the two isotopologues.

Finally, we have calculated dissociative recombination, vibrational
excitation (VE) and de-exciation (VdE) cross sections for different vibrational
levels in the entrance channels, namely for $v_i^+=1$ and $2$ and compared with
the cross sections found for $v_i^+=0$. The comparison of the DR cross sections
can be seen in figure \ref{dr_cs_vi}. We find that the DR cross section for the
ground vibrational levels exceeds significantly, by more than a factor of $2$,
those of $v_i^+=1$ and $2$, whereas the later ones have similar rates.

Figure \ref{rate_vi} shows the total DR and vibrational transition (VE, VdE)
rate coefficients for the three initial vibrational levels of the ion. We find
the highest rates for all the processes for $v^+_i=0$, their magnitude 
decreases with excitation of the initial ionic vibrational level, except for the
$\Delta v=v^+_f-v^+_i=-1$ de-excitation transitions where the $v_i^+=2$ gives
higher rates. 
 
 The trends of the observables are the same for both the cross section and rate
coefficients. Thus one can conclude that whenever the initial sample can contain
ions with excited vibrational levels, fully accounting for them will be important.
 
To aid the simple use of the 
rate coefficients shown in Fig.~\ref{rate_vi}, we have fitted them 
using  simple formulas.
The calculated DR rate coefficients for CO$^+(v=0,1,2)$ is represented by:
\begin{equation}\label{eqn:CO_DR_Interpolation}
k_{(CO^+),v}^{DR}(T_e) = A_v \, T_e^{\alpha_v} \,
\exp\left[-\sum_{i=1}^{7}\frac{B_v(i)}{i\,\,T_e^i}\right]
\end{equation}
over the electron temperature range $T_{min}<T_e<5000$~K with
$T_{min}$~=~100~K. The values given by Eq.~(\ref{eqn:CO_DR_Interpolation})
depart from the reference values by only a few percent. The parameters $A_v$,
$\alpha_v$ and $B_v(i)$ are listed in Table \ref{tab:CO_DR_Interpolation}. The
corresponding formula for the vibrational transitions (VE  and VdE) has the
form:
\begin{equation}\label{eqn:CO_VE_VdE_Interpolation}
k_{(CO^+),v'\to v''}^{VE,VdE}(T_e) = A_{v'\to v''} \, T_e^{\alpha_{v'\to v''}}
\, \exp\left[-\sum_{i=1}^{7}\frac{B_{v'\to v''}(i)}{i\,\,T_e^i}\right]
\end{equation}
over the electron temperature range $T_{min}<T_e<5000$~K. Again the
values calculated using Eq.~(\ref{eqn:CO_VE_VdE_Interpolation}) depart from the
referenced values only by a few percent. The parameters $T_{min}$, $A_{v'\to
v''}$, $\alpha_{v'\to v''}$ and the different $B_{v'\to v''}(i)$ for $i\in[1,7]$
are listed in Tables \ref{tab:CO_VE_VdE_Interpolation0} to
\ref{tab:CO_VE_VdE_Interpolation2}.

\begin{table}
\renewcommand{\arraystretch}{1.00} 
\centering
	\caption{\label{tab:CO_DR_Interpolation}Parameters used in
Eq.~(\ref{eqn:CO_DR_Interpolation}) to represent the DR rate coefficients of CO$^+$.}
  \footnotesize
	\rotatebox{90}{
    \begin{tabular}{lccccccccc}
		\hline
    $v$ &  $A_v$                      & $\alpha_v$                  &  $B_v(1)$ 
                & $B_v(2)$                    & $B_v(3)$                   &
$B_v(4)$                    & $B_v(5)$                   & $B_v(6)$             
     & $B_v(7)$ \\
		\hline
0 & .281949503$\times10^{-04}$ &  -.802579890$\times10^{00}$ &
.710130957$\times10^{03}$ & -.698308895$\times10^{06}$ &
.335177615$\times10^{09}$ & -.850620477$\times10^{11}$ &
.117104453$\times10^{14}$ &  -.826280436$\times10^{15}$ &
.233919918$\times10^{17} $ \\
1 & .715701893$\times10^{-05}$ &  -.682706983$\times10^{00}$ &
.843520677$\times10^{03}$ & -.713971932$\times10^{06}$ &
.300490272$\times10^{09}$ & -.703053442$\times10^{11}$ &
.920945293$\times10^{13}$ & -.629626035$\times10^{15}$ &
.174554736$\times10^{17}$ \\
2 & .158116294$\times10^{-05}$ & -.537454517$\times10^{00}$ &
.740102970$\times10^{03}$ & -.910764930$\times10^{06}$ &
.460910057$\times10^{09}$ & -.118698360$\times10^{12}$ &
.163967837$\times10^{14}$ & -.115670469$\times10^{16} $ &
.327041619$\times10^{17}$ \\
		\hline
    \end{tabular}}%
    \normalsize
\end{table}%

\begin{table}
\renewcommand{\arraystretch}{1.00} 
\centering
	\caption{\label{tab:CO_VE_VdE_Interpolation0}Parameters used
in Eq.~(\ref{eqn:CO_VE_VdE_Interpolation}) to represent the VE rate coefficients of CO$^+$
 ($v'$~=~0).}
  \footnotesize
	\rotatebox{90}{
    \begin{tabular}{lrrrrrrrrrr}
		\hline
    $v'\to v''$& $T_{min}$(K)     &  $A_{v'\to v''}$ & $\alpha_{v'\to v''}$     
   & $B_{v'\to v''}(1)$           & $B_{v'\to v''}(2)$           & $B_{v'\to
v''}(3)$           & $B_{v'\to v''}(4)$           & $B_{v'\to v''}(5)$          
& $B_{v'\to v''}(6)$           & $B_{v'\to v''}(7)$ \\
		\hline
0~$\to$~ 1   & 100 & .641552103$\times10^{-05}$ & -.771371374$\times10^{00}$ &
.414475403$\times10^{04}$ & -.555697606$\times10^{06}$ &
.177113555$\times10^{09}$ & -.379319033$\times10^{11}$ &
.499340108$\times10^{13}$ & -.353310994$\times10^{15}$ &
.101763661$\times10^{17}$\\
0~$\to$~ 2   & 200 & .748711072$\times10^{-05}$ & -.848613994$\times10^{00}$ &
.808028559$\times10^{04}$ & -.314102151$\times10^{07}$ &
.267788469$\times10^{10}$ & -.125332604$\times10^{13}$ &
.323013435$\times10^{15}$ & -.429933877$\times10^{17}$ &
.230779467$\times10^{19}$\\
0~$\to$~ 3   & 250 & .653782258$\times10^{-05}$ & -.889871881$\times10^{00}$ &
.100400799$\times10^{05}$ & -.153427057$\times10^{07}$ &
.166204703$\times10^{10}$ & -.968343280$\times10^{12}$ &
.307963790$\times10^{15}$ & -.505267521$\times10^{17}$ &
.334692067$\times10^{19}$\\
0~$\to$~ 4   & 300 & .410991022$\times10^{-04}$ & -.111309015$\times10^{01}$ &
.144578277$\times10^{05}$ & -.425122442$\times10^{07}$ &
.457572230$\times10^{10}$ & -.279947663$\times10^{13}$ &
.974452298$\times10^{15}$ & -.180068315$\times10^{18}$ &
.137000155$\times10^{20}$\\
0~$\to$~ 5   & 350 & .673816604$\times10^{-04}$ & -.121225593$\times10^{01}$ &
.186129714$\times10^{05}$ & -.751669966$\times10^{07}$ &
.865504846$\times10^{10}$ & -.581314625$\times10^{13}$ &
.227948002$\times10^{16}$ & -.482391689$\times10^{18}$ &
.424701037$\times10^{20}$\\
0~$\to$~ 6   & 400 & .678304142$\times10^{-04}$ & -.126655170$\times10^{01}$ &
.216968564$\times10^{05}$ & -.932715458$\times10^{07}$ &
.125788319$\times10^{11}$ & -.980812594$\times10^{13}$ &
.440562841$\times10^{16}$ & -.105632400$\times10^{19}$ &
.104556888$\times10^{21}$\\
0~$\to$~ 7   & 450 & .935214058$\times10^{-04}$ & -.135011215$\times10^{01}$ &
.242759946$\times10^{05}$ & -.866103794$\times10^{07}$ &
.131034124$\times10^{11}$ & -.117137348$\times10^{14}$ &
.603374022$\times10^{16}$ & -.165375862$\times10^{19}$ &
.186623905$\times10^{21}$\\
0~$\to$~ 8   & 500 & .127223673$\times10^{-03}$ & -.143309661$\times10^{01}$ &
.268650737$\times10^{05}$ & -.655138124$\times10^{07}$ &
.888503051$\times10^{10}$ & -.729741009$\times10^{13}$ &
.338502966$\times10^{16}$ & -.803004985$\times10^{18}$ &
.732964756$\times10^{20}$\\
0~$\to$~ 9   & 550 & .106418750$\times10^{-03}$ & -.146917756$\times10^{01}$ &
.289593294$\times10^{05}$ & -.336743387$\times10^{07}$ &
.171303281$\times10^{10}$ & .173009739$\times10^{13}$ &
-.296792983$\times10^{16}$ & .156774183$\times10^{19}$ &
-.292893172$\times10^{21}$\\
0~$\to$~ 10 & 600 & .144300838$\times10^{-04}$ & -.135396575$\times10^{01}$ &
.296834933$\times10^{05}$ & .806531715$\times10^{07}$ &
-.280264726$\times10^{11}$ & .444214735$\times10^{14}$ &
-.379357647$\times10^{17}$ & .167825913$\times10^{20}$ &
-.301076067$\times10^{22}$\\
0~$\to$~ 11 & 650 & .867766831$\times10^{-07}$ & -.958697496$\times10^{00}$ &
.257440709$\times10^{05}$ & .409137189$\times10^{08}$ &
-.119011764$\times10^{12}$ & .189804083$\times10^{15}$ &
-.170031924$\times10^{18}$ & .799092375$\times10^{20}$ &
-.152981923$\times10^{23}$\\
0~$\to$~ 12 & 700 & .325017591$\times10^{-32}$ & .446113412$\times10^{01}$ &
-.512283310$\times10^{05}$ & .493759852$\times10^{09}$ &
-.147156377$\times10^{13}$ & .246380214$\times10^{16}$ &
-.232988126$\times10^{19}$ & .115912196$\times10^{22}$ &
-.235418187$\times10^{24}$\\
		\hline
    \end{tabular}}%
    \normalsize
\end{table}%

\begin{table}
\renewcommand{\arraystretch}{1.00} 
\centering
	\caption{\label{tab:CO_VE_VdE_Interpolation1}Parameters used
in Eq.~(\ref{eqn:CO_VE_VdE_Interpolation}) to represent the VE and VdE rate coefficients
of CO$^+$  ($v'$~=~1).}
	\footnotesize
	\rotatebox{90}{
    \begin{tabular}{lrrrrrrrrrr}
		\hline
    $v'\to v''$&  $T_{min}$ & $A_{v'\to v''}$             & $\alpha_{v'\to v''}$
        & $B_{v'\to v''}(1)$           & $B_{v'\to v''}(2)$           &
$B_{v'\to v''}(3)$           & $B_{v'\to v''}(4)$           & $B_{v'\to v''}(5)$
          & $B_{v'\to v''}(6)$           & $B_{v'\to v''}(7)$ \\
		\hline
1~$\to$~0  & 100 & .633920326$\times10^{-05}$ &  -.770044378$\times10^{00}$ & 
.104178437$\times10^{04}$ &  -.551232678$\times10^{06}$ & 
.175018067$\times10^{09}$ &  -.374062491$\times10^{11}$ & 
.492117834$\times10^{13}$ &  -.348198826$\times10^{15}$ & 
.100308337$\times10^{17}$ \\
1~$\to$~2  & 100 & .112757539$\times10^{-05}$ &  -.667253557$\times10^{00}$ & 
.301658328$\times10^{04}$ &  .265128201$\times10^{06}$ & 
-.167485020$\times10^{09}$ &  .460282036$\times10^{11}$ & 
-.652022601$\times10^{13}$ &  .465696778$\times10^{15}$ & 
-.132656980$\times10^{17}$ \\
1~$\to$~3  & 150 & .562989138$\times10^{-06}$ &  -.651039420$\times10^{00}$ & 
.526482012$\times10^{04}$ &  .134397812$\times10^{07}$ & 
-.848709209$\times10^{09}$ &  .288214188$\times10^{12}$ & 
-.547449693$\times10^{14}$ &  .544047009$\times10^{16}$ & 
-.219710957$\times10^{18}$ \\
1~$\to$~4  & 200 & .130991873$\times10^{-06}$ &  -.537069718$\times10^{00}$ & 
.875368149$\times10^{04}$ &  .339794737$\times10^{06}$ & 
-.146345601$\times10^{09}$ &  .270246030$\times10^{11}$ & 
.183890443$\times10^{12}$ &  -.736617748$\times10^{15}$ & 
.699602764$\times10^{17}$ \\
1~$\to$~5  & 250 & .209439988$\times10^{-05}$ &  -.858043793$\times10^{00}$ & 
.147483446$\times10^{05}$ &  -.617683461$\times10^{07}$ & 
.628994423$\times10^{10}$ &  -.343234925$\times10^{13}$ & 
.103740710$\times10^{16}$ &  -.163563141$\times10^{18}$ & 
.104899664$\times10^{20}$ \\
1~$\to$~6  & 300 & .439087022$\times10^{-05}$ &  -.950117793$\times10^{00}$ & 
.180805100$\times10^{05}$ &  -.780708040$\times10^{07}$ & 
.975729438$\times10^{10}$ &  -.654274154$\times10^{13}$ & 
.239495154$\times10^{16}$ &  -.451595265$\times10^{18}$ & 
.342947398$\times10^{20}$ \\
1~$\to$~7  & 350 & .990587644$\times10^{-05}$ &  -.107079534$\times10^{01}$ & 
.209321916$\times10^{05}$ &  -.904877766$\times10^{07}$ & 
.124117280$\times10^{11}$ &  -.929195541$\times10^{13}$ & 
.387627434$\times10^{16}$ &  -.844294279$\times10^{18}$ & 
.746823381$\times10^{20}$ \\
1~$\to$~8  & 400 & .290364496$\times10^{-05}$ &  -.976365340$\times10^{00}$ & 
.209545755$\times10^{05}$ &  .435177075$\times10^{07}$ & 
-.136200658$\times10^{11}$ &  .175763114$\times10^{14}$ & 
-.114936867$\times10^{17}$ &  .372696383$\times10^{19}$ & 
-.475130034$\times10^{21}$ \\
1~$\to$~9  & 450 & .553151534$\times10^{-05}$ &  -.107588250$\times10^{01}$ & 
.228931552$\times10^{05}$ &  .102630980$\times10^{08}$ & 
-.292944159$\times10^{11}$ &  .393966744$\times10^{14}$ & 
-.277151718$\times10^{17}$ &  .982366990$\times10^{19}$ & 
-.138325077$\times10^{22}$ \\
1~$\to$~10& 500 & .295872921$\times10^{-18}$ &  .190367144$\times10^{01}$ & 
-.111794851$\times10^{05}$ &  .184960763$\times10^{09}$ & 
-.439761036$\times10^{12}$ &  .564818219$\times10^{15}$ & 
-.399526307$\times10^{18}$ &  .146086706$\times10^{21}$ & 
-.215320891$\times10^{23}$ \\
1~$\to$~11& 550 & .860719793$\times10^{-14}$ &  .818099637$\times10^{00}$ & 
.202216367$\times10^{04}$ &  .144200890$\times10^{09}$ & 
-.371106005$\times10^{12}$ &  .518902243$\times10^{15}$ & 
-.402326107$\times10^{18}$ &  .162115803$\times10^{21}$ & 
-.264323393$\times10^{23}$ \\
1~$\to$~12& 600 & .238429782$\times10^{-24}$ &  .309243990$\times10^{01}$ & 
-.287784203$\times10^{05}$ &  .327916264$\times10^{09}$ & 
-.895346410$\times10^{12}$ &  .135575406$\times10^{16}$ & 
-.114959696$\times10^{19}$ &  .509491570$\times10^{21}$ & 
-.917135223$\times10^{23}$ \\
1~$\to$~13& 650 & .171475246$\times10^{-53}$ &  .946512823$\times10^{01}$ & 
-.118535342$\times10^{06}$ &  .845804438$\times10^{09}$ & 
-.241202045$\times10^{13}$ &  .384508952$\times10^{16}$ & 
-.344818061$\times10^{19}$ &  .162171803$\times10^{22}$ & 
-.310615978$\times10^{24}$ \\
		\hline
    \end{tabular}}%
    \normalsize
\end{table}%

\begin{table}
\renewcommand{\arraystretch}{1.00} 
\centering
	\caption{\label{tab:CO_VE_VdE_Interpolation2}Parameters used
in Eq.~(\ref{eqn:CO_VE_VdE_Interpolation}) to represent the VE and VdE rate coefficients
of CO$^+$  ($v'$~=~2).}
	\footnotesize
	\rotatebox{90}{
    \begin{tabular}{lrrrrrrrrrr}
		\hline
    $v'\to v''$&  $T_{min}$ & $A_{v'\to v''}$             & $\alpha_{v'\to v''}$
        & $B_{v'\to v''}(1)$           & $B_{v'\to v''}(2)$           &
$B_{v'\to v''}(3)$           & $B_{v'\to v''}(4)$           & $B_{v'\to v''}(5)$
          & $B_{v'\to v''}(6)$           & $B_{v'\to v''}(7)$ \\
		\hline
2~$\to$~0  & 100 & .143264325$\times10^{-05}$ & -.676398119$\times10^{00}$ &
.770399508$\times10^{03}$ & -.649235829$\times10^{06}$ &
.280763127$\times10^{09}$ & -.687619887$\times10^{11}$ &
.943978108$\times10^{13}$ & -.672803202$\times10^{15}$ &
.193189057$\times10^{17}$ \\
2~$\to$~1  & 100 & .110150559$\times10^{-05}$ & -.664642895$\times10^{00}$ &
-.475690563$\times10^{02}$ & .274737488$\times10^{06}$ &
-.172309109$\times10^{09}$ & .472936617$\times10^{11}$ &
-.669923671$\times10^{13}$ & .478615271$\times10^{15}$ &
-.136383154$\times10^{17}$ \\
2~$\to$~3  & 100 & .296299450$\times10^{-06}$ & -.540421763$\times10^{00}$ &
.322902717$\times10^{04}$ & -.185353201$\times10^{06}$ &
.102324472$\times10^{09}$ & -.295771623$\times10^{11}$ &
.446770807$\times10^{13}$ & -.336457775$\times10^{15}$ &
.998160611$\times10^{16}$ \\
2~$\to$~4  & 150 & .278155705$\times10^{-05}$ & -.783269363$\times10^{00}$ &
.680722101$\times10^{04}$ & -.103878379$\times10^{07}$ &
.741843999$\times10^{09}$ & -.276056904$\times10^{12}$ &
.549666157$\times10^{14}$ & -.558745281$\times10^{16}$ &
.228035409$\times10^{18}$ \\
2~$\to$~5  & 200 & .266683567$\times10^{-04}$ & -.107862034$\times10^{01}$ &
.110175934$\times10^{05}$ & -.322418905$\times10^{07}$ &
.245112696$\times10^{10}$ & -.101947642$\times10^{13}$ &
.237960237$\times10^{15}$ & -.292641709$\times10^{17}$ &
.147503229$\times10^{19}$ \\
2~$\to$~6  & 250 & .542147000$\times10^{-05}$ & -.983012338$\times10^{00}$ &
.122608040$\times10^{05}$ & .321533385$\times10^{06}$ &
-.729487979$\times10^{09}$ & .465646557$\times10^{12}$ &
-.147373795$\times10^{15}$ & .236308026$\times10^{17}$ &
-.153015638$\times10^{19}$ \\
2~$\to$~7  & 300 & .317786833$\times10^{-06}$ & -.721222477$\times10^{00}$ &
.142487517$\times10^{05}$ & .102722134$\times10^{07}$ &
-.582000169$\times10^{09}$ & .484888182$\times10^{11}$ &
.644040556$\times10^{14}$ & -.223384907$\times10^{17}$ &
.220068495$\times10^{19}$ \\
2~$\to$~8  & 350 & .273643357$\times10^{-06}$ & -.744726391$\times10^{00}$ &
.184645633$\times10^{05}$ & -.445860723$\times10^{07}$ &
.783952518$\times10^{10}$ & -.677237281$\times10^{13}$ &
.310300559$\times10^{16}$ & -.723338935$\times10^{18}$ &
.674944000$\times10^{20}$ \\
2~$\to$~9  & 400 & .161814112$\times10^{-05}$ & -.947134778$\times10^{00}$ &
.222431469$\times10^{05}$ & -.163473342$\times10^{07}$ &
-.438401719$\times10^{10}$ & .980397656$\times10^{13}$ &
-.775362755$\times10^{16}$ & .276484946$\times10^{19}$ &
-.372814588$\times10^{21}$ \\
2~$\to$~10& 450 & .857949106$\times10^{-05}$ & -.113992880$\times10^{01}$ &
.247605739$\times10^{05}$ & .513853850$\times10^{07}$ &
-.233874625$\times10^{11}$ & .348626787$\times10^{14}$ &
-.254552925$\times10^{17}$ & .917457255$\times10^{19}$ &
-.130303783$\times10^{22}$ \\
2~$\to$~11& 500 & .222319239$\times10^{-11}$ & .337546441$\times10^{00}$ &
.802866148$\times10^{04}$ & .966824466$\times10^{08}$ &
-.237713634$\times10^{12}$ & .308999968$\times10^{15}$ &
-.219586955$\times10^{18}$ & .804421605$\times10^{20}$ &
-.118654044$\times10^{23}$ \\
2~$\to$~12& 550 & .967268411$\times10^{-11}$ & .141405228$\times10^{00}$ &
.108888796$\times10^{05}$ & .100995726$\times10^{09}$ &
-.264030143$\times10^{12}$ & .370381461$\times10^{15}$ &
-.287172365$\times10^{18}$ & .115625983$\times10^{21}$ &
-.188366916$\times10^{23}$ \\
2~$\to$~13& 600 & .859235026$\times10^{-11}$ & .675737504$\times10^{-01}$ &
.117098302$\times10^{05}$ & .110881879$\times10^{09}$ &
-.305681884$\times10^{12}$ & .463253664$\times10^{15}$ &
-.392591768$\times10^{18}$ & .173889056$\times10^{21}$ &
-.312870721$\times10^{23}$ \\
2~$\to$~14& 650 & .164458016$\times10^{-23}$ & .280896167$\times10^{01}$ &
-.257956296$\times10^{05}$ & .332078837$\times10^{09}$ &
-.946887885$\times10^{12}$ & .150871571$\times10^{16}$ &
-.135222037$\times10^{19}$ & .635565472$\times10^{21}$ &
-.121648562$\times10^{24}$ \\
		\hline
    \end{tabular}}%
    \normalsize
\end{table}%

\section{Conclusion}

This paper presents a theoretical study of DR of CO$^{+}$ over a broad range of
electron energies including several dissociation states with different
symmetries. Although our basic approach is
\textit{ab initio}, using such methods it is not yet possible to compute the
position of the dissociative curves involved in the
DR process accurately enough to give reliable results. We have therefore used
potential curves derived from spectroscopic data on CO to calibrate the
\textit{ab initio} data. Our DR calculations using these calibrated curves give
good agreement with the low-energy experimental data, up to about $3$ eV,
significantly improving the results of existing previous studies.
We note however the very recent study of DR N$_2^+$ performed by \cite{jtx}
based on fully {\it ab initio} potential energy curves and couplings computed
with the R-matrix method \cite{jt560,jt574}. N$_2^+$ is isoelectronic with CO$^+$,
although its higher symmetry makes it somewhat easier to treat. These studies
suggest that a similar, accurate, fully {\it ab initio} treatment of the CO$^+$ 
problem should be possible in the fairly near future.

A major motivation of this paper was to show and test the applicability of our
method, whenever there are significant lacks of reliable quantum chemistry
calculations. We have performed calculations making use of $4$ dissociation
states in three different symmetry accounting Rydberg states up to four partial
waves. The calculations were done in the highest order of complexity possible.

On overall, one can say that the agreement achieved between theory and
experiment over a significant range of energies is satisfactory, the trends being qualitatively reproduced
for this diatomic system with many-electrons. This suggests that our
calibrated \textit{ab initio} approach provides a suitable
procedure for studying  other many-electron systems for which pure \textit{ab
initio} calculations are not reliable.

The found results are of fundamental importance and very promising, suggesting
for a need of further studies especially for improved  and more accurate
potential energy curves and couplings not just for the molecular ion but for
neutral molecule as well.

\section*{Acknowledgments}

The authors thank Ch. Jungen and S. L.  Guberman for numerous and fruitful
discussions. 
They acknowledge support from the International Atomic Energy Agency via the
Coordinated Research Project "Light Element Atom, Molecule and Radical Behaviour
in the Divertor and Edge Plasma Regions," from Agence Nationale de la Recherche
via the projects "SUMOSTAI" (No. ANR-09-BLAN-020901) and "HYDRIDES" (No.
ANR-12-BS05-0011-01), from the IFRAF-Triangle de la Physique via the project
"SpecoRyd," and from the Centre National de la Recherche Scientifique via the
programs "Physique et Chimie du Milieu Interstellaire," and the PEPS projects
"Physique th\'eorique et ses interfaces" TheMS and TPCECAM. They also
acknowledge generous financial support from La R\'egion Haute-Normandie via the
CPER "THETE" project, and the GRR Electronique, Energie et Matriaux, from the
"F\'ed\'eration de Recherche Energie, Propulsion, Environnement," and from the
LabEx EMC$^3$, via the project PicoLIBS (No. ANR-10-LABX-09-01). K.C. thanks the
department Institut des Sciences de l'Ing\'enierie et des Syst\'emes (INSIS) of
CNRS for a research grant in 2013, and Laboratoire Ondes et Milieux Complexes
(LOMC) of Le Havre University for hospitality.

\section*{Data availability}
Upon a reasonable request, the data supporting this article will be provided by the corresponding author.

\end{document}